\input amstex
\documentstyle{amsppt}
\pagewidth{360pt}
\pageheight{587pt}
\accentedsymbol\tx{\tilde x}
\accentedsymbol\ty{\tilde y}
\def\Res{\operatornamewithlimits{Res}}
\topmatter
\title
On the geometry of point-expansions for certain class
of differential equations of the second order.
\endtitle
\rightheadtext{On the geometry of point-expansions  \dots}
\author
O.~N.~Mikhailov, R.~A.~Sharipov.
\endauthor
\thanks
Paper is written under financial support of European fund INTAS
(project \#93-47, coordinator of project S.I.~Pinchuk) and
Russian fund for Fundamental Researches (project \#96-01-00127,
head of project Ya.T.~Sultanaev). Work is also supported by the
grant of Academy of Sciences of the Republic Bashkortostan
(head of project N.M.~Asadullin).
\endthanks
\address
50-let SSSR str. 40-154, Ufa 450071, Russia.\newline
\endaddress
\email
yavdat\@bgua.bashkiria.su
\endemail
\address
Rabochaya str. 5, Ufa 450003, Russia.
\endaddress
\email
root\@bgua.bashkiria.su
\endemail
\abstract Second order ordinary differential equations of the form
$y''=P(x,y)+4\,Q(x,y)\,y'+6\,R(x,y)\,{y'}^2+4\,S(x,y)\,{y'}^3
+L(x,y)\,{y'}^4$ are considered and their point-expansions are
constructed. Geometrical structures connected with these expansions
are described.
\endabstract
\endtopmatter
\loadbold
\document
\head
1. Introduction.
\endhead
     Class of ordinary differential equations of the second
order with the third order polynomials in $y'$ in right hand
side
$$
y''=P(x,y)+3\,Q(x,y)\,y'+3\,R(x,y)\,{y'}^2+S(x,y)\,{y'}^3
\tag1.1
$$
is closed with respect to the point transformations of the form
$$
\cases
\tx=\tx(x,y),\\
\ty=\ty(x,y).
\endcases
\tag1.2
$$
This means that on applying this change of variables to \thetag{1.1}
we shall get another equation of the same form
$$
\ty''=\tilde P(\tx,\ty)+3\,\tilde Q(\tx,\ty)\,\ty'+3\,\tilde R(\tx,\ty)
\,{\ty'}^2+\tilde S(\tx,\ty)\,{\ty'}^3.
\tag1.3
$$
For two particular equations \thetag{1.1} and \thetag{1.3} the
problem of the existence of the point transformation \thetag{1.2}
transforming one of them into another is known as {\it the problem
equivalence}. Different aspects of this problem were studied
during the long time (see \cite{1-22}). In paper \cite{23} one
can find the scheme of point-classification for the equations
of the form \thetag{1.1} and the complete list of cases and subcases
arising in this scheme.\par
    Let's increase by 1 the degree of polynomial in right hand
side of \thetag{1.1}. Then we get the class of equations
$$
y''=P(x,y)+4\,Q(x,y)\,y'+6\,R(x,y)\,{y'}^2+4\,S(x,y)
\,{y'}^3+L(x,y)\,{y'}^4,
\hskip-2em
\tag1.4
$$
which is not closed with respect to the point transformations
\thetag{1.2}. Change of variables \thetag{1.2} gives rise to
the denominator in \thetag{1.4}:
$$
y''=\frac{P(x,y)+4\,Q(x,y)\,y'+6\,R(x,y)\,{y'}^2+4\,S(x,y)
\,{y'}^3+L(x,y)\,{y'}^4}{Y(x,y)-X(x,y)\,y'}.
\hskip-2em
\tag1.5
$$
Class of the equations \thetag{1.5} is closed with respect to
the point transformations \thetag{1.2}. Let's call it the
point-expansion for the class of the equations \thetag{1.4}.
The main purpose of this paper consists in describing the
geometrical structures connected with these equations.
\head
2. Point transformations.
\endhead
     Change of variables \thetag{1.2} in differential equations can
be treated as the change of some curvilinear coordinates on the plane
for others. This is the way for geometrization of different problems
associated with such transformations. We shall take the point
transformations \thetag{1.2} to be regular and we denote by $T$ and $S$
direct and inverse matrices of Jacoby for them:
$$
\xalignat 2
&S=\Vmatrix
x_{\sssize 1.0} &x_{\sssize 0.1}\\
\vspace{1.3ex}
y_{\sssize 1.0} &y_{\sssize 0.1}
\endVmatrix,
&&T=\Vmatrix
\tilde x_{\sssize 1.0} &\tilde x_{\sssize 0.1}\\
\vspace{1.3ex}
\tilde y_{\sssize 1.0} &\tilde y_{\sssize 0.1}
\endVmatrix.
\tag2.1
\endxalignat
$$
According to the paper \cite{23} by means of double indices in
\thetag{2.1} and in what follows we denote partial derivatives.
Thus for the function $f$ of two arguments by $f_{\sssize p.q}$
we denote the differentiation $p$ times with respect to the
first argument and $q$ times with respect to the second one.
\par
     Here is the formula for transforming the first derivative
$y'$ under the point transformations \thetag{1.2}:
$$
y'=\frac{y_{\sssize 1.0}+y_{\sssize 0.1}\,\tilde y'}
{x_{\sssize 1.0}+x_{\sssize 0.1}\,\tilde y'}.
\tag2.2
$$
Appropriate formula for the second derivative $y''$ is the
following one:
$$
\aligned
y''&=\frac{(x_{\sssize 1.0}+x_{\sssize 0.1}\,\tilde y')
(y_{\sssize 2.0}+2\,y_{\sssize 1.1}\,\tilde y'+y_{\sssize 0.2}
\,(\tilde y')^2+y_{\sssize 0.1}\,\tilde y'')}
{(x_{\sssize 1.0}+x_{\sssize 0.1}\,\tilde y')^3}-\\
\vspace{2ex}
&\qquad-\frac{(y_{\sssize 1.0}+y_{\sssize 0.1}
\,\tilde y')(x_{\sssize 2.0}+2\,x_{\sssize 1.1}\,\tilde y'+
x_{\sssize 0.2}\,(\tilde y')^2+x_{\sssize 0.1}\,\tilde y'')}
{(x_{\sssize 1.0}+x_{\sssize 0.1}\,\tilde y')^3}.
\endaligned
\tag2.3
$$
Substituting \thetag{2.2} and \thetag{2.3} into \thetag{1.5} we
can get the transformation rules for the coefficients of the
equation \thetag{1.5}. First let's consider the transformation
rules for the parameters $X$ and $Y$ in denominator:
$$
\aligned
X&=u\,\left(S^1_1\,\tilde X+S^1_2\,\tilde Y\right),\\
Y&=u\,\left(S^2_1\,\tilde X+S^2_2\,\tilde Y\right).
\endaligned
\tag2.4
$$
Here $u=u(x,y)$ is an arbitrary function of two variables.
It is present in \thetag{2.4} since numerator and denominator
of a fraction can be multiplied by the same factor $u(x,y)\neq
0$ without change of its value. This doesn't change the form
of dependence on $y'$ too. For the sake of certainty let's
take $u=1$. Then the transformation rules for $X$ and $Y$ can
be written in matrix form using the matrix $S$ from \thetag{2.1}:
$$
\Vmatrix X\\ \vspace{1ex} Y\endVmatrix
=\Vmatrix S^1_1 &S^1_2 \\ \vspace{1ex} S^2_1 &S^2_2 \endVmatrix
\cdot
\Vmatrix \tilde X\\ \vspace{0.6ex} \tilde Y\endVmatrix.
\tag2.5
$$
The rule \thetag{2.5} is exactly the rule of transformations of
the components of a vector under the change of local curvilinear
coordinates on the plane (for more details see \cite{24}). It
enables us to construct the vector field $\boldsymbol\alpha$ with
the following components:
$$
\xalignat 2
&\alpha^1=X,
&&\alpha^2=Y.
\tag2.6
\endxalignat
$$
It's clear that $\boldsymbol\alpha\neq 0$ since if both components
of this field vanish  simultaneously,  then  this  leads  to  the
vanishing
of the denominator in \thetag{1.5}.\par
     Apart from ordinary vectorial and tensorial fields, in the theory
of point transformations for the equations \thetag{1.1} and \thetag{1.5}
we have the pseudotensorial fields with certain weight (see \cite{25}
and \cite{23}).
\definition{Definition 2.1} Pseudotensorial field of the type
$(r,s)$ and weight $m$ is the multidimensional array $F^{i_1\ldots\,
i_r}_{j_1\ldots\,j_s}$ which is transformed as
$$
F^{i_1\ldots\,i_r}_{j_1\ldots\,j_s}=
(\det T)^m\sum\Sb p_1\ldots p_r\\ q_1\ldots q_s\endSb
S^{i_1}_{p_1}\ldots\,S^{i_r}_{p_r}\,\,
T^{q_1}_{j_1}\ldots\,T^{q_s}_{j_s}\,\,
\tilde F^{p_1\ldots\,p_r}_{q_1\ldots\,q_s}
$$
under the point changes of variables of the form \thetag{1.2}.
\enddefinition
Let's substitute $y'=z$ into the right hand side of \thetag{1.5}
and let's consider the resulting fraction as the rational function
of the variable $z$:
$$
f(z)=\frac{P+4\,Q\,z+6\,R\,z^2+4\,S\,z^3+L\,z^4}{Y-X\,z}.
\tag2.7
$$
Function \thetag{2.7} has a pole at the point $z_0=Y/X$. Let's
calculate the residue of this function at $z_0$ and denote by
$\Omega=\Omega(x,y)$ the following quantity:
$$
\Omega=-X^5\Res_{z=z_0}f(z).
\tag2.8
$$
It's easy to get an explicit expression for the quantity $\Omega$
defined in \thetag{2.8}:
$$
\Omega=P\,X^4+4\,Q\,X^3\,Y+6\,R\,X^2\,Y^2+4\,S\,X\,Y^3+L\,Y^4.
\tag2.9
$$
It's easy also to check that the quantity \thetag{2.9} is transformed
as the pseudoscalar field of the weight $-2$ under the point
transformations \thetag{1.2}:
$$
\Omega=(\det T)^{-2}\,\tilde\Omega.
\tag2.10
$$
The relationship \thetag{2.10} can be derived by means of direct
calculations on the base of transformation rules for the coefficients
of the equation \thetag{1.5}. As for these rules themselves, we shall
not write them here since they are rather huge in explicit form.
\head
3. Special coordinates.
\endhead
   For the further analysis of the equations \thetag{1.5} we shall
use the following well-known theorem on the straightening of the
vector field (see \cite{26} or \cite{27}).
\proclaim{Theorem 3.1} For any nonzero vector field $\boldsymbol\alpha$
on the plane with the components $\alpha^1$ and $\alpha^2$ in
coordinates $x$ and $y$ one can find the point transformation
\thetag{1.2} such that in coordinates $\tx$ and $\ty$ this
field $\boldsymbol\alpha$ is expressed by $\tilde\alpha^1=0$ and
$\tilde\alpha^2=1$.
\endproclaim
     Applying this theorem to the equation \thetag{1.5} we obtain
the following result.
\proclaim{Corollary} Any equation \thetag{1.5} can be transformed
to the form \thetag{1.4} by means of point change of variables of
the form \thetag{1.2}.
\endproclaim
     Let's choose special coordinates $x$ and $y$, in which the
vector field $\boldsymbol\alpha$ has the unit components:
$$
\xalignat 2
&\tilde\alpha^1=0,
&\tilde\alpha^2=1.
\tag3.1
\endxalignat
$$
Now let's consider the point transformations that preserve the
condition \thetag{3.1}. They form special subclass among
general point transformations of the form \thetag{1.2} given
by the following relationships:
$$
\xalignat 2
&\tilde x=h(x),
&\tilde y=y+g(x).
\tag3.2
\endxalignat
$$
By differentiating these relationships \thetag{3.2} we find the
components of transition matrices $S$ and $T$ from \thetag{2.1}:
$$
\xalignat 2
&S=\frac{1}{h'}\Vmatrix 1 & 0\\ \vspace{1.3ex}-g'& h'\endVmatrix,
&&T=\Vmatrix h' & 0\\ \vspace{1.3ex}g'& 1\endVmatrix.
\tag3.3
\endxalignat
$$
Since these matrices are not degenerate, we get $\det T=h'(x)\neq 0$.
\par
     In the special coordinates, which were chosen above, the
equation \thetag{1.5} has the form \thetag{1.4}. This is due to
\thetag{3.1}. Let's consider the transformation rules for the
parameters $L$ and $S$ in \thetag{1.4} under the point
transformations \thetag{3.2}, which preserve the condition
\thetag{3.1}:
$$
\xalignat 2
&L=\frac{1}{{h'}^2}\,\tilde L,
&&S=\frac{g'}{{h'}^2}\,\tilde L+\frac{1}{h'}\,\tilde S.
\tag3.4
\endxalignat
$$
First of the relationships \thetag{3.4} means that the parameter
$L$ is transformed as pseudoscalar field of the weight $-2$.
This is not surprising since in case, when the conditions
\thetag{3.1} hold, the formula \thetag{2.9} for the field
$\Omega$ gives $\Omega=L$.\par
    {\bf Conclusion}. Parameter $L$ in special coordinates
defines the pseudoscalar field of the weight $-2$ that can
be extended as the field $\Omega$ for arbitrary coordinates.
\par
    Let's take one more vector field $\boldsymbol\psi$ other
than vector field $\boldsymbol\alpha$. Its components in
special coordinates are given by the following relationships:
$$
\xalignat 2
&\psi^1=L,
&\psi^2=-S.
\tag3.5
\endxalignat
$$
Then we can represent the formulas \thetag{3.4} in matrix form:
$$
\Vmatrix \psi^1\\ \vspace{1ex}\psi^2 \endVmatrix
=\frac{1}{h'}\Vmatrix S^1_1 &S^1_2 \\ \vspace{1ex} S^2_1 &S^2_2
\endVmatrix\cdot
\Vmatrix \tilde\psi^1\\ \vspace{0.6ex} \tilde\psi^2\endVmatrix.
\tag3.6
$$
From \thetag{3.6} we see that the field $\boldsymbol\psi$ defined
by \thetag{3.5} in special coordinates is the pseudovectorial field
of the weight $-1$. Note that the condition of noncollinearity of
the fields $\boldsymbol\alpha$ and $\boldsymbol\psi$ coincides
with $L\neq 0$ in special coordinates, and with $\Omega\neq 0$ in
arbitrary ones. For the equations \thetag{1.4} and \thetag{1.5}
this condition surely holds since had it been broken, this would
reduce the equations \thetag{1.4} and \thetag{1.5} to the form
\thetag{1.1}.\par
     Having defined the field $\boldsymbol\psi$ in special coordinates
it's natural to calculate its components in arbitrary coordinates.
Let's denote them by $M$ and $N$:
$$
\xalignat 2
&\psi^1=M,
&\psi^2=N.
\tag3.7
\endxalignat
$$
Then $M$ and $N$ are defined by the following two formulas:
$$
\gather
\aligned
 M=Q\,X^3&+ 3\,R\,X^2\,Y + 3\,S\,X\,Y^2 + L\,Y^3 + \\
 &+\frac{X}{4}\left(X^2\,Y_{\sssize 1.0}
 -X\,X_{\sssize 1.0}\,Y
 +X\,Y\,Y_{\sssize 0.1}
 -X_{\sssize 0.1}\,Y^2\right).\hskip-2em
\endaligned
\tag3.8\\
\aligned
 N&=-P\,X^3 - 3\,Q\,X^2\,Y - 3\,R\,X\,Y^2 - S\,Y^3 + \\
 &+\frac{Y}{4}\left(X^2\,Y_{\sssize 1.0}
 -X\,X_{\sssize 1.0}\,Y
 +X\,Y\,Y_{\sssize 0.1}
 -X_{\sssize 0.1}\,Y^2\right).
\endaligned
\tag3.9
\endgather
$$
In order to prove the formulas \thetag{3.8} and \thetag{3.9} first
let's check that in special coordinates they give $M=L$ and $N=-S$.
Then we should verify the following transformation rules for $M$
and $N$ given by \thetag{3.8} and \thetag{3.9}:
$$
\Vmatrix M\\ \vspace{1ex}N \endVmatrix
=(\det T)^{-1}\Vmatrix S^1_1 &S^1_2 \\ \vspace{1ex} S^2_1 &S^2_2
\endVmatrix\cdot
\Vmatrix \tilde M\\ \vspace{0.6ex} \tilde N\endVmatrix.
\tag3.10
$$
This can be done by direct calculations based on the transformation
rules for the coefficients of the equation \thetag{1.5}.\par
     Vector field $\boldsymbol\alpha$, pseudovectorial field
$\boldsymbol\psi$, and pseudoscalar field $\Omega$ are bound
with the following relationship:
$$
\Omega=\sum^2_{i=0}\sum^2_{j=0}d_{ij}\,\psi^i\,\alpha^j.
\tag3.11
$$
Here $d_{ij}$ is the unit skew-symmetric matrix $2\times 2$:
$$
d_{ij}=\Vmatrix\format \r&\quad\l\\ 0 & 1\\-1 & 0\endVmatrix.
\tag3.12
$$
Matrix \thetag{3.12} defines twice-covariant pseudotensorial
field of the weight $-1$.\par
     After explicit calculation of the sums in right hand side of
\thetag{3.11} we may rewrite this formula as follows:
$$
\Omega=M\,Y-N\,X.
\tag3.13
$$
Now the relationship \thetag{3.13} can be easily derived from
\thetag{2.9}, \thetag{3.8}, and \thetag{3.9} by direct calculations.
\par
\head
4. Associated equation.
\endhead
     Now let's use the components of the fields $\boldsymbol\alpha$
and $\boldsymbol\psi$, and the pseudoscalar field $\Omega$ for to
construct the following fraction:
$$
k(z)=\frac{(N-M\,z)^4}{\Omega^3(Y-X\,z)}.
\tag4.1
$$
Both functions, $k(z)$ from \thetag{4.1} and $f(z)$ from \thetag{2.7},
have simple pole at the same point $z_0=Y/X$. Due to the relationships
\thetag{3.13} their residues are equal. Therefore the difference of
these two functions is a polynomial:
$$
f(z)-k(z)=P^*(x,y)+3\,Q^*(x,y)\,z+3\,R^*(x,y)\,z^2+S^*(x,y)\,z^3.
\tag4.2
$$
It's easy to calculate the coefficients of the polynomial \thetag{4.2}
in special coordinates, where $X=0$ and $Y=1$:
$$
\xalignat 2
&P^*=P-\frac{S^4}{L^3},
&&Q^*=\frac{4Q}{3}-\frac{4S^3}{L^2},\\
\vspace{-1.2ex}
&&&\tag4.3\\
\vspace{-1.2ex}
&R^*=2R-\frac{2S^2}{L},
&&S^*=0.
\endxalignat
$$
As for the coefficients $P^*$, $Q^*$, $R^*$, $S^*$ in arbitrary
coordinates, they can be expressed trough $P$, $Q$, $R$, $S$, $L$,
$X$, $Y$. But these expressions are huge enough. Instead of writing
them in explicit form, we shall prove the following theorem.
\proclaim{Theorem 4.2} Coefficients of the polynomial \thetag{4.2}
are defined uniquely by the differential equation \thetag{1.5}.
\endproclaim
\demo{Proof} Using \thetag{4.1} and \thetag{4.2} we can calculate
$P^*$, $Q^*$, $R^*$, $S^*$ trough $P$, $Q$, $R$, $S$, $L$, $X$, and
$Y$. However, the parameters $P$, $Q$, $R$, $S$, $L$, $X$, $Y$ are
not determined uniquely by the equation \thetag{1.5}. Their choice
admits gauge arbitrariness of the form
$$
\xalignat 2
&P\to\varphi(x,y)\,P,&&Q\to\varphi(x,y)\,Q,\\
&R\to\varphi(x,y)\,R,&&S\to\varphi(x,y)\,S,\\
\vspace{-1.5ex}
&&&\tag4.4\\
\vspace{-1.5ex}
&L\to\varphi(x,y)\,L,&&X\to\varphi(x,y)\,X,\\
&Y\to\varphi(x,y)\,Y
\endxalignat
$$
since numerator and denominator of the fraction in \thetag{1.5}
can be multiplied by the same factor $\varphi(x,y)$. Substituting
\thetag{4.4} into the formulas \thetag{3.8} and \thetag{3.9}
we get gauge transformations for $M$ and $N$:
$$
\xalignat 2
&M\to\varphi(x,y)^4\,M,&&N\to\varphi(x,y)^4\,N.
\tag4.5
\endxalignat
$$
Now let's substitute \thetag{4.4} and \thetag{4.5} into \thetag{3.13}.
This gives gauge transformations for the field $\Omega$:
$$
\Omega\to\varphi(x,y)^5\,\Omega.
\tag4.6
$$
From \thetag{4.4}, \thetag{4.5}, and \thetag{4.6}, on substituting them
into \thetag{4.1}, we find that the function $k(z)$ is invariant under
gauge transformations \thetag{4.4}. Hence polynomial \thetag{4.2} is
also invariant under such transformations. Theorem~4.1 is proved.
\qed\enddemo
Now we shall derive the transformation rules for $P^*$, $Q^*$, $R^*$,
and $S^*$ under the point transformations \thetag{1.2}. Let's
substitute $z=y'$ into the fraction \thetag{4.1} and let's apply
the formula \thetag{2.2} for the derivative $y'$. Taking into
account the transformation rules \thetag{2.5}, \thetag{2.10}, and
\thetag{3.10} we get
$$
k(y')=
\frac{\det S}{(x_{\sssize 1.0}+x_{\sssize 0.1}\,\tilde y')^3}\,
\tilde k(\tilde y').
\tag4.7
$$
Here $\tilde k(z)$ is a fraction of the form \thetag{4.1} with
$X$, $Y$, $M$, $N$, and $\Omega$ replaced by $\tilde X$, $\tilde Y$,
$\tilde M$, $\tilde N$, and $\tilde \Omega$. In order to get
transformation rules for $P^*$, $Q^*$, $R^*$, $S^*$ let's write
the equation \thetag{1.5} as follows:
$$
y''=P^*(x,y)+3\,Q^*(x,y)\,y'+3\,R^*(x,y)\,{y'}^2+S^*(x,y)\,{y'}^3+
k(y').
\tag4.8
$$
Now we substitute \thetag{2.2} and \thetag{2.3} into \thetag{4.8}
and take into account \thetag{4.7}. This leads us to the
following central result of our paper.
\proclaim{Theorem 4.2} Coefficients $P^*(x,y)$, $Q^*(x,y)$,
$R^*(x,y)$, and $S^*(x,y)$ in polynomial \thetag{4.2} are
transformed exactly as the coefficients of the differential
equation
$$
y''=P^*(x,y)+3\,Q^*(x,y)\,y'+3\,R^*(x,y)\,{y'}^2+S^*(x,y)\,{y'}^3
\tag4.9
$$
under the point transformations of the form \thetag{1.2}.
\endproclaim
     This equation \thetag{4.9} is called {\it canonically
associated} equation for \thetag{1.5}. It is of the form \thetag{1.1}.
From theorems~4.1 and 4.2 we get the following obvious result.
\proclaim{Theorem 4.3} If two equations of the form \thetag{1.5}
are point-equivalent, i.\,e. one of them is obtained from another
by some point transformation \thetag{1.2}, then their associated
equations \thetag{4.9} are also point-equivalent.
\endproclaim
\head
5. Acknowledgments.
\endhead
     Authors are grateful to Professors E.G.~Neufeld, V.V.~Sokolov,
A.V.~Bocharov, V.E.~Adler, N.~Kamran, V.S.~Dryuma, A.B.~Sukhov,
and M.V.~Pavlov for the information, for worth advises, and for
help in finding many references below.\par

\enddocument
\end